\documentclass[hyper]{JHEP} 

\usepackage{epsfig}

\newcommand\fverb{\setbox\pippobox=\hbox\bgroup\verb}

\newcommand\fverbdo{\egroup\medskip\noindent%

            \fbox{\unhbox\pippobox}\ }

\newcommand\fverbit{\egroup\item[\fbox{\unhbox\pippobox}]}

\newbox\pippobox

\title{New $p+1$ Dimensional Non-relativistic Theories from
Euclidean Stable and Unstable
Dp-Branes}
\author{J. Kluso\v{n}\\
Department of
Theoretical Physics and Astrophysics\\
Faculty of Science, Masaryk University\\
Kotl\'{a}\v{r}sk\'{a} 2, 611 37, Brno\\
Czech Republic\\
E-mail: \email{klu@physics.muni.cz}}
\preprint{\hepth{0905.1483}}

 \abstract{In this paper we continue the
study of non-relativistic $p+1$
dimensional theories that we started
at arXiv:0904.1343. We
extend the analysis presented there
 to the case of
stable and unstable Dp-branes.}

\def\mH{\mathcal{H}}

\def\tY{\tilde{Y}}
\def\iD{(D^{-1})}
\def\tmJ{\tilde{\mathcal{J}}}

\def\mJ{\mathcal{J}}
\def\bG{\mathbf{G}}

\def\tA{\tilde{A}}

\def\bx{\mathbf{x}}
\def\by{\mathbf{y}}
\def \mD{\mathcal{D}}

\newcommand{\tmD}{\tilde{\mathcal{D}}}
\newcommand{\tN}{\tilde{N}}

\newcommand{\bQ}{\mathbf{Q}}

\newcommand{\hA}{\hat{A}}
\newcommand{\mF}{\mathcal{F}}

\newcommand{\hQ}{\hat{Q}}

\newcommand{\hPi}{\hat{\Pi}}

\newcommand{\mG}{\mathcal{G}}

\def\ket #1{\left|#1\right>}

\def \bAi{\left(\mathbf{A}^{-1}\right)}

\def \bA{\mathbf{A}}

\newcommand{\bT}{\mathbf{T}}
\newcommand{\ba}{\mathbf{a}}

\newcommand{\mL}{\mathcal{L}}

\def\bai{(\mathbf{a}^{-1})}

\def \tY{\tilde{Y}}
\def\pb #1{\left\{#1\right\}}

\begin{document}
\section{Introduction}\label{first}
In past few  months
  P. Ho\v{r}ava formulated  new
interesting  theories with anisotropic
scaling between time and spatial
dimensions in series of papers
\cite{Horava:2009if,Horava:2009uw,Horava:2008ih,Horava:2008jf}
\footnote{For recent study of
cosmological aspects and further
conceptual issues of these theories,
see
\cite{Myung:2009va,
Ghodsi:2009rv,Cai:2009qs,Rama:2009px,Kehagias:2009is,Nishioka:2009iq,
Gao:2009er,Orlando:2009en,Cai:2009dx,Myung:2009dc,
Mukohyama:2009zs,Chen:2009ka,Sotiriou:2009gy,Colgain:2009fe,Gao:2009bx,
Piao:2009ax,Volovik:2009av,Cai:2009ar,
Cai:2009pe,Nastase:2009nk,Izawa:2009ne,Brandenberger:2009yt,
Mukohyama:2009gg,Lu:2009em,Kiritsis:2009sh,Calcagni:2009ar,Takahashi:2009wc,
Jenkins:2009un,Visser:2009fg}.}.
 This phenomenon is well known
from the study of condensed matter
systems at quantum criticality
\cite{Ardonne:2003wa}. The similar
systems  have  been investigated   from
the point of view of non-relativistic
form of AdS/CFT correspondence
\cite{Blau:2009gd,Volovich:2009yh,Pal:2009np,
Son:2008ye,Balasubramanian:2008dm,Goldberger:2008vg,
Barbon:2008bg,Wen:2008hi,Herzog:2008wg,Maldacena:2008wh,
Adams:2008wt,Minic:2008xa,Chen:2008ad,Colgain:2009wm,Bagchi:2009my,
Alishahiha:2009hg,Donos:2009en,Pal:2009yp,Danielsson:2008gi,
Taylor:2008tg,Adams:2008zk,Akhavan:2008ep,Rangamani:2008gi,Schvellinger:2008bf,
Sachdev:2008ba,Hartnoll:2008rs,Lin:2008pi,Yamada:2008if,Duval:2008jg,
Kovtun:2008qy,Kachru:2008yh}
 \footnote{For another approach to the
 study of
non-relativistic systems in string
theory,see for example
\cite{Lee:2009mm,Galajinsky:2009dw,Nakayama:2009cz,
Nakayama:2008qm,Gomis:2004pw,Brugues:2004an,Kluson:2006xi,
Gomis:2006wu,Gomis:2006xw,Sakaguchi:2006pg,Gomis:2005bj,Gomis:2005pg}.}.

The construction of the theories with
anisotropic scaling is based on the
following question: Is it possible to
find a  $p+1$ dimensional quantum
theory such that its ground state wave
functional reproduces the partition
function of $p$ dimensional theory?

One particular example of such
$p$ dimensional theories was studied in
our previous paper \cite{Kluson:2009sm}
where we constructed new
non-relativistic $p+1$ dimensional
theories from the Nambu-Gotto action
for p-branes in general background. The
goal of this paper is to implement
 similar ideas to the case when
$p$ dimensional system is either stable
or unstable D(p-1)-brane action. We
also presume that  this action is
embedded in general background with
non-trivial dilaton, gravity and NSNS
two-form field. As in case of p-brane
theory we should stress that this $p$
dimensional action is highly non-linear
with all well known consequences for
the renormazibility and quantum
analysis of given action. Further, even
if we consider Dp-brane actions that
are well known from superstring
theories we do not address the question
how these new theories  can be realized
in superstring  ones. In particular, we
do not worry about relation between
dimensions of Dp-branes and their
realizations in either IIA or type IIB
theories.

Even if we leave the question of
explicit  realization of these theories
in the framework of string theory open
we mean that these  new theories are
interesting in own sense. These $p+1$
dimensional theories are invariant
under spatial diffeomorphism and
spatial gauge transformations as well
under global time translation. Then
following \cite{Horava:2008ih} we show
that these symmetries can be extended
to the space-time diffeomorphism that
respect the preferred codimension-one
foliation of $p+1$ dimensional
space-time by the slices at fixed time
known as \emph{ foliation-preserving
diffeomorphism}. Note that these
transformations consist a space-time
dependent spatial diffeomorphism
together with time-dependent
reparameterization of time. Further, we
also extend the spatial dependent gauge
transformations to the space-time
dependent ones. Then in order to
achieve of the invariance of  $p+1$
dimensional  action under these
symmetries  we introduce  gauge fields
$N^i$ and $N$ to maintain its
invariance under foliation preserving
diffeomorphism and $A_t$ to maintain
gauge invariance. As in case of
non-critical p-brane theory
\cite{Kluson:2009sm}  these new gauge
fields will be crucial for the correct
Hamiltonian formulation of the theory
as the theory of constraint systems
and we  show that the Hamiltonian
can be expressed  as a linear combination
of constraints.

We also address the question of the tachyon
condensation on the world-volume of $p+1$
dimensional theory. We show that this theory
has a natural spatial dependent solution known
as a tachyon kink and we argue that the
resulting codimension one defect corresponds
to stable D(p-1)-brane at criticality.

The organization of this paper is as
follows.  In the next section
(\ref{second}) we formulate the
non-critical $p+1$ dimensional theories
that obey detailed balance condition in
the sense that their potential term is
proportional to the variation of
D(p-1)-brane action.   In section
(\ref{third}) we generalize the gauge
symmetries of this $p+1$ dimensional
theory when we extend rigid time
translation and spatial diffeomorphism
to the foliation-preserving
diffeomorphism and spatial dependent
gauge symmetry to the space-time
dependent one. In section
(\ref{fourth}) we develop the
Hamiltonian formalism for given theory
and we calculate the algebra of
constraints. In section  (\ref{fifth})
we study the tachyon condensation on
the world-volume of $p+1$ dimensional
Dp-brane that contains tachyon and we
argue that it leads to the emergence of
stable $p$ dimensional D(p-1)-brane
theory at criticality.
 Finally in section
(\ref{sixth}) we outline our results
and suggest possible extension of this
work.
\section{Stable and Unstable D-branes at
Criticality}\label{second} In this
section we formulate $p+1$ dimensional
Dp-brane theories that obey detailed
balance conditions. We closely
 follow \cite{Kluson:2009sm}.

Let us consider Euclidean DBI action
for unstable D(p-1)-brane that is embedded
in general $10$-dimensional background
\begin{eqnarray}\label{actW}
W&=&\tau_{p-1} \int d^{p}\bx
 e^{-\Phi}V(T)
\sqrt{\det\bA} \ , \nonumber \\
\bA_{ij}&=&\partial_i
Y^M\partial_j Y^N
g_{MN}+2\pi\alpha'\mF_{ij}+
2\pi\alpha'
\partial_i T\partial_j T  \ , \nonumber \\
  \mF_{ij}&=&
\partial_i A_j-\partial_j A_i-
(2\pi\alpha')^{-1}b_{MN}\partial_i
Y^M\partial_j Y^N
 \ , \nonumber \\
\end{eqnarray}
where
$\tau_{p-1}=\frac{\sqrt{2}}{(2\pi)^{p-1}
\alpha'^{\frac{p}{2}}}$ is an unstable
D(p-1)-brane tension,
$\bx=(x^1,\dots,x^p)$ are the
world-volume coordinates, $Y^M,
M,N=1,\dots,10$ are world-volume
scalars that parameterize the embedding
of the D(p-1)-brane in target
space-time. Further, $T$ is
world-volume tachyon mode with the
potential $V(T)$ that is even function
of $T$ with the property
\begin{equation}
V(T\rightarrow \pm \infty)=0 \ , \quad
V(0)=1 \ .
\end{equation}
Finally,  $A_i$ is a world-volume gauge
field and  $g_{MN},b_{MN}$ and $\Phi$
are background metric, NS-NS two-form
and dilaton respectively. Note that the
stable D(p-1)-brane can be derived from
(\ref{actW}) be striping the tachyon
contribution ($V(T)=1 \ , T=0$) and
replacing $\tau_{p-1}$ with
$T_{p-1}=\frac{1}{\sqrt{2}}\tau_{p-1}$.
For simplicity we consider the
background with zero Ramond-Ramond
fields even if the analysis can be
easily generalized to this case as
well.

Let us now discuss symmetries of the action
(\ref{actW}).
By construction this action   is invariant
under  world-volume diffeomorphism:
\begin{equation}\label{spd}
x'^i=x'^i(\bx) \
\end{equation}
under which the world-volume fields
transform as
\begin{eqnarray}
Y'^M(\bx')=Y^M(\bx) \ , \quad
T'(\bx')=T(\bx) \ , \quad
A_i(\bx')=A'_j(\bx)(D^{-1})^j_i \ ,
 \nonumber
\\
\end{eqnarray}
where we introduced $p\times p$ matrix
$D^i_j=\frac{\partial x'^i}{\partial
x^j}$ that is generally function of
$\bx$. Then it is easy to show that
$G_{ij}$ and $B_{ij}$ defined as
\begin{equation}
G_{ij}(\bx)=g_{MN}(Y(\bx))\partial_iY^M(\bx)\partial_j
Y^N(\bx) \ , \quad
B_{ij}(\bx)=b_{MN}(Y(\bx))\partial_i Y^M(\bx)\partial_j
Y^N(\bx)
\end{equation}
together with $F_{ij}=\partial_iA_j-\partial_j
A_i$
transform under (\ref{spd}) as
\begin{eqnarray}
 G'_{ij}(\bx')&=&
G_{kl}(\bx)\iD^k_i
\iD^l_j \ , \nonumber \\
B'_{ij}(\bx')&=&B_{kl}(\bx)
\iD^k_i \iD^l_j \ ,
\nonumber \\
\mF'_{ij}(\bx')&=&
\mF_{kl}(\bx)\iD^k_i  \iD^j_l
\nonumber \\
\end{eqnarray}
and consequently
\begin{equation}
\sqrt{\det \bA'(\bx')}=\frac{1}{|\det
D|}\sqrt{\det \bA(\bx)} \ .
\end{equation}
Then  using the fact that the element
$d^p\bx$ transforms as $d^p\bx'=|\det
D|d^p\bx$  under (\ref{spd}) we find
that
\begin{equation}
d^p\bx' \sqrt{\det \bA'(\bx')}= d^p\bx
\sqrt{\det \bA(\bx)} \ .  \nonumber \\
\end{equation}
It can be also easily shown that
the action (\ref{actW}) is invariant under
local gauge transformations:
\begin{eqnarray}
A'_i(\bx)=A_i(\bx)+\partial_i f(\bx) \ ,
\quad
Y'^M(\bx)=Y^M(\bx) \ .
\nonumber \\
\end{eqnarray}
Now we are ready to construct $p+1$
dimensional theory. The starting point
is the presumption that
configuration space of $p+1$ dimensional
 theory is
spanned by $Y^M(\bx),T(\bx)$ and
$A_i(\bx)$. In other words
 the states of this theory in
Schr\"{o}dinger representation are
functionals of $Y^M(\bx),T(\bx)$ and
$A_i(\bx)$. Explicitly, we presume that
the $p+1$ dimensional quantum  theory
contains the operators $\hat{T}(\bx),
\hat{Y}^M(\bx)$ and $\hat{A}_i(\bx)$
together with their conjugate momenta $
\hPi_T(\bx),\hPi_M(\bx)$ and
$\hPi^i(\bx)$ with standard commutation
relations
\begin{eqnarray}\label{comre}
\left[\hat{Y}^M(\bx),\hPi_N(\by)\right]&=&
i\delta^M_N\delta(\bx-\by) \ ,
\nonumber \\
\left[\hat{T}(\bx),\hPi_T(\by)\right]&=&
i\delta(\bx-\by) \ ,  \nonumber \\
\left[\hat{A}_i(\bx),\hPi^j(\by)\right]&=&
i\delta^j_i\delta(\bx-\by) \  . \nonumber \\
\end{eqnarray}
Further, the eigenstate  of the
operators $
\hat{Y}^M(\bx),\hat{T}(\bx)$ and
$\hat{A}_i(\bx)$ is the state
$\ket{Y(\bx),T(\bx),A(\bx)}$ defined as
\begin{eqnarray}
\hat{Y}^M(\bx)\ket{Y(\bx),T(\bx),A(\bx)}&=&
Y^M(\bx)\ket{Y(\bx),T(\bx),A(\bx)} \ , \nonumber \\
\hat{T}(\bx)\ket{Y(\bx),T(\bx),A(\bx)}&=&
T(\bx)\ket{Y(\bx),T(\bx),A(\bx)} \ , \nonumber \\
\hat{A}_i(\bx)\ket{Y(\bx),T(\bx),A(\bx)}&=&
A_i(\bx)\ket{Y(\bx),T(\bx),A(\bx)} \ .
\nonumber \\
\end{eqnarray}
In the  Schr\"{o}dinger representation
the state $\ket{\Psi}$ of the system is
represented by wave functional
$\Psi[Y(\bx),T(\bx),A(\bx)]$ and the
action of the operators
$\hat{Y}^M(\bx), \hat{T}(\bx)$ and
$\hat{A}_i(\bx)$ on this state is
represented by multiplication with
$Y^M(\bx),T(\bx)$ and $A_i(\bx)$
respectively. Then the commutation
relations  (\ref{comre}) suggest that
the momentum operators have following
functional form in the Schr\"{o}dinger
representation:
\begin{equation}
\hPi_M(\bx)=-i\frac{\delta}{\delta Y^M(\bx)}  \ ,
\quad
\hPi_T(\bx)=-i\frac{\delta}{\delta T(\bx)} \ ,
\quad
\hPi^i(\bx)=-i\frac{\delta}{\delta A_i(\bx)}
\ .
\end{equation}
As the next step in the construction of
$p+1$ dimensional quantum theory we determine
corresponding Hamiltonian. This construction
is based on two presumptions. The first one
is an existence of the   wave functional  $\Psi[Y(\bx),T(\bx),A(\bx)]$
that has the form
\begin{equation}
\Psi[Y(\bx),T(\bx),A(\bx)]=
\exp(-W) \ ,
\end{equation}
where $W$ is defined in
(\ref{actW}).
The second one is the definition
of the operators  $\hat{Q}_M(\bx),
\hat{Q}_T(\bx),\hat{Q}^i(\bx)$
\begin{eqnarray}\label{defhQ}
\hQ_M(\bx)&=&-i\hat{\Pi}_M(\bx)+
\frac{1}{2}\frac{\delta \hat{W}}{\delta
\hat{Y}^M(\bx)} \ , \nonumber \\
\hQ_T(\bx)&=&-i\hPi_T(\bx)
+\frac{1}{2}\frac{\delta \hat{W}}{\delta
\hat{T}(\bx)} \ , \nonumber \\
\hQ^i(\bx)&=&-i\hPi^i(\bx)
+\frac{1}{2}\frac{\delta \hat{W}}{\delta
\hA_i(\bx)} \ . \nonumber \\
\end{eqnarray}
Then we finally  presume that  the
quantum mechanical Hamiltonian density of $p+1$
dimensional theory takes the form
\begin{eqnarray}\label{QMh}
\hat{\mathcal{H}}=
\frac{\kappa^2}{2}
(\hat{Q}_M\hat{\mG}^{MN}\hat{Q}_N+
\hat{Q}_T \hat{\mG}^{TT}\hat{Q}_T+\hat{Q}^i
\hat{\mG}_{ij}\hat{Q}^j) \ ,
\nonumber \\
\end{eqnarray}
where $\kappa$ is the coupling constant, and
where $\hat{\mG}^{MN},\hat{\mG}^{TT}$ and $\hat{\mG}^{ij}$
that generally depend on
of $\hat{Y}^M,\hat{T},\hat{A}_i$  will be defined bellow
when we will discuss the classical Lagrangian form of
the theory.
 Note
that in the Schr\"{o}dinger representation
the operators  defined in
(\ref{defhQ})
take the form
\begin{eqnarray}
\hat{Q}_M(\bx)&=&
\frac{\delta}{\delta Y^M(\bx)}+
\frac{1}{2}\frac{\delta W}{\delta Y^M(\bx)} \ ,
\nonumber \\
\hat{Q}_T(\bx)&=&\frac{\delta}{\delta T(\bx)}
+\frac{1}{2}
\frac{\delta W}{\delta T(\bx)} \ , \nonumber \\
\hat{Q}^i(\bx)&=&\frac{\delta}{\delta A_i(\bx)}
+\frac{1}{2}
\frac{\delta W}{\delta A_i(\bx)} \
\nonumber \\
\end{eqnarray}
and it is clear that these operators annihilate
the wave functional $\Psi[Y(\bx),A(\bx),T(\bx)]=
\exp(-W)$ and consequently this state is
eigenstate of the Hamiltonian with zero energy.
However it is crucial that this is only formal
construction. For example, we do not discuss
the issue  whether this state is normalizable.
We should rather consider this construction
as  the motivation for  the particular  form
of the quantum Hamiltonian density
(\ref{QMh}) that implies corresponding form
of the classical Hamiltonian density. In fact,
in what follows we restrict to
the classical  description when we
 replace operators with their classical
analogues. At the classical level we
also   specify the form of the matrices
$\mG$ that appear in (\ref{QMh}) and we
suggest that they have the  form
\begin{eqnarray}\label{defMG}
\mG^{MN}&=&\frac{2\pi\alpha'g^{MN}}
{\tau_{p-1}e^{-\Phi}V(T)\sqrt{\det\bA}} \ , \nonumber \\
\mG^{TT}&=&\frac{1}{\tau_{p-1}
e^{-\Phi}V(T)
\sqrt{\det\bA}} \ , \nonumber \\
\mG_{ij}&=&\frac{\bA^S_{ij}}{2\pi\alpha'\tau_{p-1}e^{-\Phi}
V(T)\sqrt{\det\bA}} \ , \nonumber \\
\end{eqnarray}
where we defined symmetric and anti-symmetric
part of the matrix $\bA_{ij}$ as
\begin{equation}
\bA_{ij}^S=\frac{1}{2}(\bA_{ij}+\bA_{ji}) \ ,
\quad
\bA_{ij}^A=\frac{1}{2}(\bA_{ij}-\bA_{ji}) \ .
\end{equation}
Let us say few comments that explain
our choice of the matrices
(\ref{defMG}).
First of all it was shown
in \cite{Seiberg:1999vs} that
the natural open string metric
on the world-volume of Dp-brane is given
by combination $G_{ij}+\mF_{ij}$. Further,
 it was
also argued in many papers
\cite{Kutasov:2003er,Sen:1999md,Garousi:2000tr,Bergshoeff:2000dq,
Kluson:2000iy,Sen:2003tm}
\footnote{For review and extensive list
of references, see \cite{Sen:2004nf}.}
 that
 the tachyon kinetic term should appear
in the linear combination with $G_{ij}$
 and $F_{ij}$. These arguments explain our
 choice of $\bA_{ij}^S$. Note also that
presence of the  factor
$\tau_{p-1}V(T)$ in (\ref{defMG}) is
dictated by requirement that the
Lagrangian of
 non-relativistic $p+1$ dimensional theory
 contains
 an overall factor $\tau_{p-1}V(T)$
that is crucial for the correct interpretation of
the tachyon kink on the world-volume of
$p+1$ dimensional theory.
Finally, since we want the Lagrangian
density with the factor $e^{-\phi}$ we
had to include the factor $e^{\phi}$ into
the definition of the Hamiltonian density. In
summary, we argue that the classical Hamiltonian
density takes the form
\begin{eqnarray}\label{mhclas}
\mH&=&\frac{\kappa^2}{2}
\left(i\Pi_M(\bx)+\frac{1}{2}\frac{\delta W}{
\delta Y^M(\bx)}\right)
\frac{2\pi\alpha'g^{MN}(\bx)}
{\tau_{p-1}e^{-\phi(\bx)}V(T)\sqrt{\det \bA(\bx)}}
\left(-i\Pi_N(\bx)+\frac{1}{2}\frac{\delta W}
{\delta Y^N(\bx)}\right)+\nonumber \\
&+&\frac{\kappa^2}{2}
\left(i\Pi_T(\bx)+\frac{1}{2}\frac{\delta W}
{\delta T(\bx)}\right)
\frac{1}{\tau_{p-1}e^{-\phi(\bx)}V(T)\sqrt{\det \bA(\bx)}}
\left(-i\Pi_T(\bx)+\frac{1}{2}
\frac{\delta W}{\delta T(\bx)}\right)+\nonumber \\
&+&\frac{\kappa^2}{2}
\left(i\Pi^i(\bx)+\frac{1}{2}
\frac{\delta W}{\delta A_i(\bx)}\right)
\frac{(\bA^S)_{ij}(\bx)}{2\pi\alpha'\tau_{p-1}e^{-\phi(\bx)}
V(T)
\sqrt{\det \bA(\bx)}}
\left(-i\Pi^j(\bx)+\frac{1}{2}
\frac{\delta W}{\delta A_j(\bx)}\right) \ .
\nonumber \\
\end{eqnarray}
Now we determine corresponding Lagrangian
density. Using the equation of motion for
$Y^M,T,A_i$ we find
\begin{eqnarray}
\partial_t Y^M(\bx)&=&\pb{Y^M(\bx),H}=
\kappa^2 \frac{2\pi\alpha'g^{MN}(\bx)}
{\tau_{p-1}e^{-\phi(\bx)}V(T)\sqrt{\det\bA(\bx)}}\Pi_N(\bx) \ ,
\nonumber \\
\partial_t T(\bx)&=&
\pb{T(\bx),H}=
\kappa^2 \frac{1}{\tau_{p-1}e^{-\phi(\bx)}V(T)
\sqrt{\det\bA(\bx)}}\Pi_T(\bx) \ , \nonumber \\
\partial_t A_i(\bx)&=&
\pb{A_i(\bx),H}=
\kappa^2 \frac{(\bA^S)_{ij}(\bx)}{2\pi\alpha'\tau_{p-1}
e^{-\phi(\bx)}
V(T)\sqrt{\det \bA}}\Pi^j(\bx)
\nonumber \\
\end{eqnarray}
and hence the
Lagrangian density takes the form
\begin{eqnarray}\label{mLD}
\mL &=& \partial_tY^M\Pi_M+
\partial_t T\Pi_T+
\partial_t A_i\Pi^i-\mH=\mL_K+\mL_P \ ,
\nonumber  \\
\mL_K&=&\frac{\tau_{p-1}}{2\kappa^2}e^{-\phi}
V(T)\sqrt{\det \bA}\frac{1}{2\pi\alpha'}
\partial_t Y^Mg_{MN}\partial_tY^N+
\nonumber \\
&=&\frac{\tau_{p-1}}{2\kappa^2}e^{-\phi}
V(T)\sqrt{\det\bA}\partial_t T\partial_t T+
\frac{2\pi\alpha'\tau_{p-1}}{2\kappa^2}e^{-\phi}V(T)\sqrt{\det\bA}
\partial_t A_i \bAi^{ij}_S A_j \ ,
\nonumber \\
\mL_P &=&-\frac{\kappa^2\tau_{p-1}}{8}
e^{-\phi}V(T)\sqrt{\det\bA}
[2\pi\alpha'\mJ_Mg^{MN}\mJ_N+
\mJ_T \mJ_T+(2\pi\alpha')^{-1}\mJ^i \bA^S_{ij}\mJ^j] \ ,
\nonumber \\
\end{eqnarray}
where we defined $\mJ_M,\mJ_T $ and $\mJ^i$
as
\begin{eqnarray}
\frac{\delta W}{\delta T(\bx)}
&=&\tau_{p-1}e^{-\phi}V\sqrt{\det \bA}\left
[\frac{V'}{V}-\frac{2\pi\alpha'}{e^{-\phi}\sqrt{
\det \bA}}\partial_i[e^{-\phi}V\partial_jT
\bAi^{ji}_S\sqrt{\det\bA}]\right]\nonumber \\
&\equiv&
\tau_{p-1}e^{-\phi}V(T)\sqrt{\det \bA}\mJ_T(\bx) \ ,
\nonumber \\
\frac{\delta W}{\delta Y^M(\bx)}&=&
\tau_{p-1}e^{-\phi}V(T)\sqrt{\det\bA}
\left[\frac{\partial_M [e^{-\phi}]}{e^{-\phi}}
+\frac{1}{2}(\partial_M G_{ij}-\partial_M B_{ij})
\bAi^{ji}-\right.\nonumber \\
&-&\frac{1}{e^{-\phi}V(T)\sqrt{\det\bA}}
\partial_i[e^{-\phi}V(T)g_{MN}\partial_j Y^N \bAi^{ji}_S
\sqrt{\det\bA}]+\nonumber \\
&+&\left.\frac{1}{e^{-\phi}V(T)\sqrt{\det\bA}}
\partial_i[e^{-\phi}V(T)b_{MN}\partial_j Y^N
\bAi^{ji}_A]\right]\equiv
\nonumber \\
&\equiv &\tau_{p-1}e^{-\phi}V(T)\sqrt{\det\bA}\mJ_M(\bx) \ ,
\nonumber \\
\frac{\delta W}{\delta A_i(\bx)}&=&
-\tau_{p-1}e^{-\phi}V(T)\sqrt{\det \bA}\times
\nonumber \\
&\times&
\frac{2\pi\alpha'}{e^{-\phi}V(T)\sqrt{\det\bA}}
\partial_j[e^{-\phi}V(T)
\bAi^{ij}\sqrt{\det\bA}]\equiv
\nonumber \\
&\equiv & \tau_{p-1}e^{-\phi}V(T)
\sqrt{\det \bA}\mJ^i(\bx) \
\nonumber \\
\end{eqnarray}
that transform
under spatial diffeomorphism (\ref{spd}) as
\begin{eqnarray}
\mJ'_M(\bx')=\mJ_M(\bx) \ ,
\quad
\mJ'_T(\bx')=\mJ_T(\bx) \ ,
\quad
\mJ'^i(\bx')=\mJ^{j}(\bx)D_j^{ \ i}(\bx) \ .
\nonumber \\
\end{eqnarray}
Using these results we consequently find
\begin{equation}
\mL_K(\bx',t')=\frac{1}{\det D(\bx)}
\mL_K(\bx,t) \ , \quad
\mL_P(\bx',t')=\frac{1}{\det D(\bx)}
\mL_P(\bx,t)
\end{equation}
that implies that the action $S=\int dt
d^p\bx \mL$ is invariant under spatial
diffeomorphism (\ref{spd}). Further, it
can be also easily shown that the
action is invariant under \emph{global
time translation}
\begin{equation}\label{deltatcon}
t'=t+\delta t \ , \quad  \delta t=\mathrm{const} \ ,
\end{equation}
where the world-volume modes transform
as
\begin{equation}
Y'^M(t',\bx)=Y^M(t,\bx) \ , \quad
T'(t',\bx)=T(t,\bx) \ , \quad
A'_i(t',\bx)=A_i(t,\bx) \ .
\end{equation}
Finally we determine scaling dimension
of world-volume modes and coupling
constants \footnote{Since we consider
the classical theory only these
dimensions are classical "engineering"
dimensions.}. To do this let us presume
following scaling
\begin{eqnarray}\label{scald}
x'^i&=&\lambda^{-1}x^i \ , \quad t'=\lambda^{-z}t \ ,
\nonumber \\
 Y'^M(\bx')&=&
\lambda^{-1}Y(\bx) \ , \quad   T'(\bx')=
T(\bx) \ , \quad  A'_i(\bx')=\lambda A_i(\bx) \ ,
\nonumber \\
(\alpha')'&=&\lambda^{-2}\alpha' \ , \quad
\tau'_{p-1}=\lambda^{p}\tau_{p-1} \ , \quad
\kappa=\lambda^{[\kappa]}\kappa \  ,
\nonumber \\
\end{eqnarray}
where $\lambda$ is constant scaling parameter.
Let us say few words about this  scaling.  First
of all it is natural to presume that $x$ have
standard scaling dimension $[x]=-1$. Then
the anisotropic nature of the theory
is reflected by the scaling dimension of $t$ that
generally is different from $x'$s.
 Further, since
$Y^M$'s describe the embedding of the D(p-1)-brane into
the target space-time it is again natural to
presume that it  scales as in (\ref{scald}). In other
words $Y^M$'s have the same dimensions
as in D-brane theory. Then we can also
demand   $T$ and
$A_i$ have the standard scaling dimensions
as in  D-brane theory (\ref{scald}).
Finally the scaling dimension of $\alpha'$
and $\tau_{p-1}$ is obvious as it follows
from their string theory definition.

Let us now discuss the consequence of
the scaling (\ref{scald}). First of all
it implies that
\begin{equation}
\bA'_{ij}(\bx')=\bA_{ij}(\bx)
\ , \quad
\tau_{p-1}'d^p\bx'
=\tau_{p-1}d^p\bx
\end{equation}
and hence we find that $W$ is invariant
under scaling while the currents $\mJ$
scale as
\begin{eqnarray}
\mJ'_M(\bx')=\lambda \mJ_M(\bx) \ ,
\quad
\mJ'_T(\bx')=\mJ_T(\bx) \ , \quad
\mJ'^i(\bx)=\lambda^{-1}\mJ^i(\bx) \ .
\nonumber \\
\end{eqnarray}
As a consequence we find that the
Lagrangian densities
 (\ref{mLD}) scale as
\begin{eqnarray}
\mL'_P(\bx')&=&\lambda^{p+2[\kappa]}\mL_P(\bx) \ ,
\quad \nonumber \\
\mL'_K(\bx')&=&\lambda^{p-2[\kappa]+2z}\mL_P(\bx) \ .
\nonumber \\
\end{eqnarray}
Then the requirement of the invariance
of the action implies the relation
between dimension of $\kappa$ and $z$
\begin{equation}\label{kappaz}
[\kappa]=\frac{z}{2} \ .
\end{equation}
However it is clear from the form of
the Lagrangian density (\ref{mLD})
 that the effective
coupling constant is
\begin{equation}
\frac{1}{G^2}=\frac{\tau_{p-1}}{\kappa^2}
\end{equation}
that is dimensionless for
\begin{equation}
2[\kappa]-p=0
\end{equation}
that using (\ref{kappaz}) implies
the famous relation between scaling dimension
of $z$ and dimension of the theory
\begin{equation}
z-p=0 \ .
\end{equation}
However we should stress that the theory
still contain dimensionfull coupling
$\tau_p$.
On the other hand the scaling dimensions
of coupling constants are important
at the quantum analysis of the theory while
our treatment is pure classical.
\section{Extension of the symmetries}
\label{third} In this section we extend
the symmetries of the non-relativistic
action introduced above. Recall that
this action is invariant under  global
time translations (\ref{deltatcon}) and
under local spatial diffeomorphism
(\ref{spd}). Further, the action is
also invariant
 under spatial dependent
gauge transformation with parameter
 $\epsilon(\bx)$. Following
the logic given in \cite{Horava:2008ih}
we would like to extend this spatial
diffeomorphism and rigid time
translation to  \emph{the
 foliation preserving
diffeomorphism} defined as
\begin{eqnarray}\label{fpd}
\delta t=t'-t=f(t) \ , \quad
 \delta x^i=x'^i-x^i=
 \zeta^i(\bx,t) \ .
\end{eqnarray}
Note that under these transformations
the modes $Y^M,T$ transform as
\begin{eqnarray}\label{Ytr}
\partial_{t'} Y'^M(\bx',t')&=&
\partial_t Y^M(\bx,t)-\partial_t Y^M(\bx,t)\dot{f}(t)
-\partial_i Y^M(\bx,t)\partial_t \zeta^i(\bx,t) \ ,
\nonumber \\
\partial_{t'}T'(\bx',t')&=&
\partial_t T(\bx,t)-\partial_t T(\bx,t)\dot{f}
-\partial_i T(\bx,t)\partial_t \zeta^i(\bx,t) \ ,
\nonumber \\
 \partial_{x'^i}Y'^M(\bx',t')&=&
\partial_i Y^M(\bx,t)-\partial_j Y^M(\bx,t)\partial_i \zeta^j(\bx,t) \ ,
\nonumber \\
\partial_{x'^i}T'^M(\bx',t')&=&
\partial_i T(\bx,t)-\partial_j T(\bx,t)\partial_i \zeta^j(\bx,t) \ ,
\nonumber \\
\end{eqnarray}
where $\dot{f}=\frac{d f}{dt}$.
Before we proceed to the determination of the transformation
rules for $A_i$ we  extend the spatial dependent
 gauge symmetry with parameter $\epsilon(\bx)$
to the gauge symmetry that is time dependent as well so that
\begin{equation}\label{gai}
A'_i(\bx,t)=A_i(\bx,t)+\partial_i \epsilon(\bx,t) \ .
\end{equation}
However then the requirement of the
invariance of the action under time
dependent gauge transformation implies
that we have to introduce the gauge
field $A_t(t,\bx)$ that under gauge
transformation transforms as
\begin{equation}\label{gat}
A'_t(\bx,t)=A_t(\bx,t)+\partial_t \epsilon(\bx,t) \ .
\end{equation}
Using this gauge field we replace $\partial_t A_i$ with
the object $E_i$ defined as
\begin{eqnarray}
E_i(\bx,t)=\partial_t A_i(\bx,t)-\partial_i A_t(\bx,t)
\end{eqnarray}
that is invariant under the time dependent
gauge transformations (\ref{gai}) and
(\ref{gat}).
As the next step we determine how the
vectors $A_i(\bx,t),A_t(\bx,t)$ transform under
(\ref{fpd}). Obviously it is natural
to demand that they  transform as
 vectors
\begin{eqnarray}\label{trAi}
A'_i(\bx',t)
&=&A_i(\bx,t)-A_j(\bx,t)\partial_i \zeta^j(\bx,t) \ ,
\nonumber \\
A'_t(\bx',t')
&=&A_t(\bx,t)-A_t(\bx,t) \dot{f}(t)-A_j(\bx,t)\partial_t
\zeta^j(\bx,t)  \nonumber \\
\end{eqnarray}
and consequently
\begin{eqnarray}\label{Ftr}
F'_{ij}(\bx',t')
=F_{ij}(\bx,t)-F_{ik}(\bx,t)\partial_j \zeta^k(\bx,t)
-\partial_i \zeta^k (\bx,t) F_{kj}(\bx,t) \ .
\nonumber \\
\end{eqnarray}
Then using (\ref{Ytr}) and (\ref{Ftr})
we easily find that
\begin{equation}
d^p\bx' \sqrt{\det \bA'(\bx',t')}=
d^p\bx \sqrt{\det \bA(\bx,t)} \ .
\end{equation}
Further, using (\ref{trAi}) we
determine the transformation
property of $E_i$
\begin{eqnarray}\label{trEi}
 E'_i(\bx',t')=
 E_i(\bx,t)-E_i(\bx,t)\dot{f}(t)
 -E_j(\bx,t)\partial_i\zeta^j(\bx,t)
 -F_{ji}(\bx,t)\dot{\zeta}^j(\bx,t) \ .
\nonumber \\
\end{eqnarray}
Alternatively, we can determine the
transformation rules of $A_i,A_t$
from the relativistic transformation law
for $p+1$ dimensional vector $A_i,A_0$, following
the analysis performed in
\cite{Horava:2008ih}. Explicitly, let us
consider covariant $p+1$ dimensional vector
$A_\mu=(A_i,A_0)$ with corresponding
field strength $F_{\mu\nu}$
\begin{equation}
F_{ij}=\partial_i A_j-\partial_j A_i \ ,
\quad F_{0i}=\partial_0 A_i-\partial_i A_0 \ ,
\end{equation}
where $x^0=ct$. Since it is a covariant
object it transforms under general diffeomorphism
as
\begin{eqnarray}\label{trcov}
F'_{\mu\nu}(\bx',ct')=
F_{\kappa\rho}(\bx,ct)\frac{\partial x^\kappa}{\partial x'^\mu}
\frac{\partial x^\rho}{\partial x'^\nu} \ .
\nonumber \\
\end{eqnarray}
It is a simple task to show that (\ref{trcov})
imply the transformation rules
for $F_{ij}$ given (\ref{Ftr}). In order
to determine the transformation of $E_i$
note that we can write
$F_{0i}$ as
\begin{equation}
F_{0i}=\partial_0 A_i-\partial_i A_0=
\frac{1}{c}\partial_t A_i-\frac{1}{c}
\partial_i A_t=
\frac{1}{c}F_{ti} \ ,
\end{equation}
where due to the fact that $x^0=ct$ we
replaced $A_0$ with $A_0=\frac{1}{c}A_t$ holding
$A_t$ fixed in the limit $c\rightarrow \infty$. Then
it is easy to see that when we take
$\mu=0,\nu=i$ in (\ref{trcov})
and consider the limit  $c\rightarrow \infty$ we find
\begin{eqnarray}
F'_{0i}(\bx',t')&=&
\frac{1}{c}F'_{ti}(\bx',t')=
\nonumber \\
&=&
\frac{1}{c}F_{ti}(\bx,t)
-\frac{1}{c}F_{tj}(\bx,t)\partial_i
\zeta^j(\bx,t)
-\frac{1}{c}F_{ti}(\bx,t)\dot{f}(t)
-\frac{1}{c}F_{ji}(\bx,t)\dot{\zeta}^j(\bx,t) \
\nonumber \\
\end{eqnarray}
that reproduces (\ref{trEi}).

As we argued above the invariant
spatial volume element is $d^p\bx \sqrt{\det\bA}$.
It is also easy to see that
under folliation preserving
diffeomorphism the matrix
$\bAi$ transforms as
\begin{eqnarray}
\bAi'^{ij}(\bx',t')=
A^{ij}(\bx,t)+\partial_ l\zeta^i(\bx,t)
\bAi^{lj}(\bx,t)+\bAi^{il}(\bx,t)\partial_l \zeta^j
(\bx,t) \
\nonumber \\
\end{eqnarray}
and hence its symmetric form transforms
in the same way as spatial components
of $p+1$ dimensional metric. On the
other hand in order to have kinetic
part of the Lagrangian invariant under
foliation preserving diffeomorphism we
have to introduce two gauge fields
$N(\bx,t)$ and $N^i(t)$ that transform
under (\ref{fpd}) as
\begin{eqnarray}\label{trNi}
N'^i(\bx',t')&=&N^i(\bx,t)+N^j(\bx,t)\partial_j\zeta^i(\bx,t)
-N^i(\bx,t)\dot{f}(t)-\dot{\zeta}^i(\bx,t) \ , \nonumber \\
N'(t')&=& N(t)-N(t)\dot{f} \
\nonumber \\
\end{eqnarray}
and replace  $\partial_t Y^M,
\partial_t T$ and $E_i$ with
\begin{eqnarray}
\partial_t T &\rightarrow &\frac{1}{N}
(\partial_t T-N^i\partial_i T)\equiv \mD_t T \ ,
\nonumber \\
\partial_t Y^M &\rightarrow &
\frac{1}{N}(\partial_t Y^M-N^i
\partial_i Y^M)\equiv \mD_t Y^M \ ,   \nonumber \\
E_i &\rightarrow &
\frac{1}{N}(E_i-N^kF_{ki})
\equiv \mathcal{D}_t A_i \ . \nonumber \\
\end{eqnarray}
Then using (\ref{Ytr}),(\ref{Ftr}),(\ref{trEi}) and
(\ref{trNi}) we find that
\begin{eqnarray}
& &\mD_{t'}Y'^M(\bx',t')=
\mD_t Y^M(\bx,t)-\mD_jY^M(\bx,t)\partial_i\zeta^j(\bx,t) \ ,
\nonumber \\
& &\mD_{t'}T'(\bx',t')=
\mD_tT(\bx,t)-\mD_j T(\bx,t)\partial_i \zeta^j(\bx,t) \ ,
\nonumber \\
& & \mathcal{D}_{t'} A'_i(\bx',t')=
\mathcal{D}_t A_i(\bx,t)-
\mathcal{D}_t A_j(\bx,t)\partial_i\zeta^j(\bx,t)
 \ .
\nonumber \\
\end{eqnarray}
If we also take into account the fact that $N'(t')dt'=
Ndt$ we obtain that the action for non-relativistic
$p+1$-brane theory
that is  invariant under
foliation preserving diffeomorphism
takes the form
\begin{equation}\label{Sdi}
S=\int dt d^p\bx
N  \mL \ ,
\end{equation}
where
\begin{eqnarray}\label{Ldi}
\mL&=&\mL_K+\mL_P \ , \nonumber \\
 \mL_K&=&
\frac{\tau_{p-1}}{2\kappa^2}e^{-\phi}
V(T)\sqrt{\det \bA}\left[\frac{1}{2\pi\alpha'}
\mD_t Y^M
g_{MN}\mD_t Y^N+
\mD_t T\mD_tT+
2\pi\alpha'\mD_t A_i \bAi^{ij}_S
\mD_t A_j \right]
\ , \nonumber \\
\mL_P&=&-\frac{\kappa^2\tau_{p-1}}{8}
e^{-\phi}V(T)\sqrt{\det\bA}
[2\pi\alpha'\mJ_Mg^{MN}\mJ_N+
\mJ_T \mJ_T+(2\pi\alpha')^{-1}\mJ^i \bA^S_{ij}\mJ^j] \ .
\nonumber \\
\end{eqnarray}
In the next section we determine
the Hamiltonian formalism that follows
from the Lagrangian density
(\ref{Ldi}).
\section{Hamiltonian analysis}\label{fourth}
To begin with we introduce
the momenta  $\Pi_M,\Pi_T,\Pi^i,
\Pi^0,\pi_i,\pi_N$ that
are  conjugate to
$Y^M,A_i,T,A_t$ and $N^i,N$
and that have following non-zero
Poisson brackets
\begin{eqnarray}
\pb{Y^M(\bx),\Pi_N(\by)}&=&\delta^M_N
\delta(\bx-\by) \ ,
\quad \pb{T(\bx),\Pi_T(\by)}=
\delta(\bx-\by) \ , \quad
\nonumber \\
\pb{A_i(\bx),\Pi^j(\by)}&=&\delta_i^j
\delta(\bx-\by) \ ,
\quad
\pb{A_0(\bx),\Pi^0(\by)}=\delta(\bx-\by) \ ,
\nonumber \\
\pb{N^i(\bx),\pi_j(\by)}&=&\delta^i_j
\delta(\bx-\by) \ , \quad
\pb{N,\pi_N}=1 \ .
\nonumber \\
\end{eqnarray}
First of all since the Lagrangian
density (\ref{Ldi})
does not contain time derivative of
$N^i,N$ and $A_0$ the conjugate
momenta form a primary constraints
of the theory
\begin{equation}\label{picon}
\pi_i(\bx)\approx 0 \ , \quad \pi_N\approx 0 \ ,
\quad \Pi^0(\bx)\approx 0 \ .
\end{equation}
Further, the   momenta conjugate
to $Y_M,T$ and $A_i$ takes the form
\begin{eqnarray}
P_M(t,\bx)&=&\frac{\tau_{p-1}}{2\pi\alpha'\kappa^2 }
e^{-\phi}V(T)\sqrt{\det\bA}g_{MN}
\mD_tY^N \ ,
\nonumber \\
P_T(t,\bx)&=&\frac{\tau_{p-1}}{2\pi\alpha'\kappa^2 }
e^{-\phi}V(T)\sqrt{\det\bA}
\mD_t T \ , \nonumber \\
\Pi_i(t,\bx)&=&
\frac{2\pi\alpha'\tau_{p-1}}{\kappa^2}e^{-\phi}V(T)
\sqrt{\det\bA}\bAi^{ij}_S \mD_t A_j \ .
\nonumber \\
\end{eqnarray}
Then it is simple task to find
corresponding Hamiltonian density
from  (\ref{Ldi})
\begin{eqnarray}\label{Hden}
\mH&=&\partial_t Y^M \Pi_M+\partial_t T\Pi_T+
\partial_t A_i\Pi^i-\mL=
\nonumber \\
&=&N\left[\frac{\kappa^2 }
{2\tau_{p-1} e^{-\phi}V(T)\sqrt{\det\bA}}\left(
2\pi\alpha'\Pi_M g^{MN}\Pi_N+
 2\pi\alpha'\Pi_T \Pi_T+
\frac{1}{2\pi\alpha'}\Pi^i \bA_{ij}^S \Pi^j\right)
+\right. \nonumber \\
& &\left.+\frac{\kappa^2\tau_{p-1}}{8}
e^{-\phi}V(T)\sqrt{\det\bA}
[2\pi\alpha'\mJ_Mg^{MN}\mJ_N+
\mJ_T \mJ_T+(2\pi\alpha')^{-1}\mJ^i (\bA_S)_{ij}\mJ^j]
\right]+\nonumber\\
&+& N^i(\partial_i Y^M\Pi_M+
\partial_i T\Pi_T+F_{ij}\Pi^j)+
\partial_i A_0\Pi^i=
 \nonumber \\
 &=&NT_0+N^i T_i+\partial_iA_0 \Pi^i \ .
 \nonumber \\
\end{eqnarray}
The primary constraints (\ref{picon})
have to be preserved during the time evolution
of the system. In other words we require
\begin{equation}
\partial_t \pi_i(\bx)=\pb{\pi(\bx),H}\approx 0 \ ,
\quad
\partial_t \pi_N=\pb{\pi,H} \approx 0
\ , \quad
\partial_t \Pi^0(\bx)=\pb{\Pi^0(\bx),H}
\approx 0 \ .
\end{equation}
Then using the canonical Poisson
brackets and the form of the
Hamiltonian density (\ref{Hden}) we
obtain that the theory should be
supplemented with secondary constrains
in the form
\begin{equation}
T=\int d^p\bx T_0(\bx)\approx 0 \ , \quad
T_i(\bx)\approx 0 \ .
\end{equation}
and finally the consistency of the
constraint $\Pi^0\approx 0$ with
its time evolution implies
the constraint
\begin{equation}
G(\bx)\equiv
\partial_i \Pi^i(\bx)=0 \ .
\end{equation}
On the other hand it turns out
\cite{Horava:2008ih} that instead to
impose the constraint $T\approx  0$ it
is much more convenient to introduce
equivalent set of constraints $Q$ using
the fact that $T_0$ can be written as
\begin{eqnarray}
T_0&=&\frac{\kappa^2}{2}
\left[Q_M^\dag \frac{2\pi\alpha'g^{MN}}{\tau_{p-1}
e^{-\phi}V(T)
\sqrt{\det\bA}} Q_N+\right.\nonumber \\
&& \left. +Q_T^\dag \frac{2\pi\alpha'}{\tau_{p-1}
e^{-\phi}V(T)
\sqrt{\det\bA}}Q_T
+Q_i \frac{\bAi_S^{ij}}{2\pi\alpha'\tau_{p-1}
e^{-\phi}V(T)
\sqrt{\det\bA}}Q_j\right] \ ,
\nonumber \\
\end{eqnarray}
where
\begin{eqnarray}
Q_M&=&-i\Pi_M
+\frac{\tau_{p-1}}{2}e^{-\phi}V(T)\sqrt{\det\bA}
\mJ_M \ , \nonumber \\
Q_T&=&-i \Pi_T+\frac{\tau_{p-1}}{2}
e^{-\phi}V(T)\sqrt{\det\bA}\mJ_T \ ,  \nonumber \\
Q^i&=&-i\Pi^i+\frac{\tau_{p-1}}{2}
e^{-\phi}V(T)\sqrt{\det \bA}\mJ^i \ .
\nonumber \\
\end{eqnarray}
Let us now introduce
the  smeared form of
the constraints defined as
\begin{eqnarray}\label{smcon}
\bT(\zeta)&=&\int d^p\bx
\zeta^i(\bx)T_i(\bx) \ , \nonumber \\
\bQ_Y(\Lambda_Y)&=&
\int d^p \bx
\Lambda^M_Y(\bx)Q_M(\bx) \ , \quad
\bQ_T(\Lambda_T)=
\int d^p \bx
\Lambda_T(\bx)Q_T(\bx) \ , \nonumber \\
\nonumber \\
\bG(\epsilon)&=&
\int d^p\bx
\epsilon(\bx)G(\bx) \ ,
\quad
\bQ_A(\Lambda^A)=
\int d^p \bx
\Lambda^A_i(\bx)Q^i(\bx) \  \quad
\nonumber \\
\end{eqnarray}
and determine their algebra.
First of all we find
\begin{eqnarray}
\pb{\bG(\epsilon),A_i(\bx)}&=&
\partial_i \epsilon(\bx) \ ,
\nonumber \\
\pb{\bG(\epsilon),F_{ij}(\bx)}&=&
0 \ .
\nonumber \\
\end{eqnarray}
Then using the fact that $\mJ'$s and $\bA$ are
invariant under gauge transformations
we find that
\begin{eqnarray}\label{pbGQ}
\pb{\bG(\epsilon),\bQ(\Lambda)}=0
\ , \quad
\pb{\bG(\epsilon),\bT(\zeta)}=0  \ .
\nonumber \\
\end{eqnarray}
Further, it is straightforward exercise
to determine the Poisson brackets of
$\bT(\zeta)$ with $\bT(\xi)$
\begin{eqnarray}\label{pbTT}
\pb{\bT(\zeta),\bT(\xi)}=
\bT(\zeta^j\partial_j \xi^i-\xi^j\partial_j\zeta^i)
+\bG(-2\zeta^iF_{ik}\xi^k) \ .
\nonumber \\
\end{eqnarray}
Let us now consider the Poisson
brackets that contain $Q$'s. It
can be easily shown, following
\cite{Kluson:2009sm}
 that the Poisson brackets of
two $Q$'s  vanish
\begin{eqnarray}\label{pbQMN}
\pb{Q_M(\bx),Q_N(\by)}&=&
\pb{Q_M(\bx),Q_T(\by)}=
\pb{Q_M(\bx),Q^i(\by)}=0 \ ,
\nonumber \\
\pb{Q_T(\bx),Q_T(\by)}&=&
\pb{Q_T(\bx),Q^i(\bx)}=
\pb{Q^i(\bx),Q^j(\by)}=0 \ .
\nonumber \\
\end{eqnarray}
In order to determine
 the Poisson
bracket between $\bT(\zeta)$ and
$Q_M,Q_T$ we again follow
 \cite{Kluson:2009sm} and we find
\begin{eqnarray}\label{Q,T}
 \pb{\bT(\zeta),
Q_M(\bx)}
=-\partial_i Q_M(\bx)\zeta^i(\bx)-
Q_M(\bx)\partial_i\zeta^i(\bx) \   \nonumber \\
\end{eqnarray}
and
\begin{eqnarray}
\pb{\bT(\zeta),Q_T(\bx)}
=-
\partial_i Q_T(\bx)\zeta^i(\bx)-
Q_T(\bx)\partial_i\zeta^i(\bx)\ .
\nonumber \\
\end{eqnarray}
On the other hand the calculation
of the Poisson bracket between $\bT(\zeta)$
and $Q^i(\bx)$ is more involved however
after some effort  we find
\begin{eqnarray}\label{TzQi}
\pb{\bT(\zeta),Q^i(\bx)}=-\partial_k
\zeta^k(\bx)Q^i(\bx)-\zeta^k(\bx)
\partial_k Q^i(\bx)+\partial_k
\zeta^i(\bx)Q^k(\bx) +\zeta^i(\bx)
G(\bx)  \ .
\nonumber \\
\end{eqnarray}
Then finally using these
results we find that the
   smeared form of these
Poisson bracket takes the form
\begin{eqnarray}\label{pbsmeared}
\pb{\bT(\zeta),\bQ_Y(\Lambda_Y)}&=&
=\bQ_Y(\partial_i\Lambda_Y\zeta^i)
\ , \nonumber \\
\pb{\bT(\zeta),\bQ_T(\Lambda_T)}&=&
\bQ_T(\partial_i\Lambda_T \zeta^i) \ ,
\nonumber \\
\pb{\bT(\zeta),\bQ_A(\Lambda^A)}&=&
\bG(\Lambda^A_i\zeta^i)+
\bQ_A(\partial_k\Lambda^A_i
\zeta^k+\Lambda^A_k\partial_i\zeta^k) \ .
\nonumber \\
\end{eqnarray}
These results imply that the
requirement that the   set of
constraints $\bT, \bQ_Y,\bQ_T,\bQ_A$
and $\pi_i,\pi_N,\Pi^0$ is preserved
during the time evolution of the system
 does not
generate additional ones.
 More precisely, note that
the total Hamiltonian $H_T$ can be written
as the sum of smeared constraints defined
in (\ref{smcon})
\begin{eqnarray}
H_T&=&\int d^p\bx \mH_T=\nonumber \\
&=&
\frac{\kappa^2}{2}\left(
\bQ_Y\left(N\frac{2\pi\alpha'
Q^\dag_N g^{NM}}{\tau_{p-1}e^{-\phi}V(T)
\sqrt{\det\bA}}\right)+
\bQ_T\left(N\frac{2\pi\alpha'
Q^\dag_T }{\tau_{p-1}e^{-\phi}V(T)
\sqrt{\det\bA}}\right)+\right.
\nonumber \\
&& \left.+
\bQ_A\left(N\frac{Q^\dag_j \bAi^{ji}_S}
{\tau_{p-1}2\pi\alpha'e^{-\phi}
V\sqrt{\det\bA}}\right)
+\bT(N^i)+\bG(-A_0)\right) \ ,
\nonumber \\
\end{eqnarray}
where the expression in the parenthesis
are parameters of corresponding
constraints. Then using the form of the
algebra of constraints written above we
find that their  time evolution vanish
on constraint surface and consequently
these constraints do not generate
additional ones.  In other words the
full set of constraints of the theory
is
\begin{eqnarray}
& &\pi_i(\bx) \ , \quad \pi_N \ , \quad
\Pi^0(\bx) , \nonumber \\
& &\bT(\zeta) \ , \quad
\bQ_Y(\Lambda_Y) \ , \quad
\bQ_T(\Lambda_T) \ , \quad
 \bQ_A(\Lambda^A) \ , \quad
\bG(\epsilon) \ . \nonumber \\
\end{eqnarray}
Then   in the same
way as in \cite{Kluson:2009sm}
we can formally proceed to the quantum
formulation of the theory
 where
we replace constraint functions
by corresponding operators
$\hat{\bT}(\zeta),\hat{\bQ}_Y(\Lambda_Y),
\hat{\bQ}_T(\Lambda_T),
\hat{\bQ}_A(\Lambda^A)$ and $\hat{\bG}(\epsilon)$.
Further we can introduce the
 wave functional
\begin{equation}
\Psi[Y(\bx),A(\bx),T(\bx)]=\exp\left(-W\right)
\end{equation}
that is clearly annihilated by all
constraints functions and at the same
time it is a  state of the
Hamiltonian $H_T$ with zero eigenvalue.
However since we are interested in the
classical form of the theory we are not
going to proceed in this direction
further.
\section{Tachyon condensation}\label{fifth}
In this section we will study the
spatial dependent tachyon condensation on an unstable
non-relativistic Dp-brane with
the Lagrangian density given in (\ref{Ldi}).
 Our goal is to show that
the tachyon condensation in the
form of kink leads to D(p-1)-brane
exactly in the same way as in
superstring theory
\cite{Sen:2003tm,Kluson:2005fj}.
However due to the complexity of
the non-relativistic Lagrangian density
(\ref{Ldi})
 we restrict ourselves
to the case of flat background where
$g_{MN}=\delta_{MN} \ , e^{\Phi}=e^{\phi_0}=1
 \ , b_{MN}=0$.

To begin with note that Lagrangian
density for stable non-relativistic
D(p-1)-brane takes the form
\begin{equation}\label{Dp-1}
S=\int dt d^{p-1}\bx
N  \mL^{st} \ ,
\end{equation}
where
\begin{eqnarray}
\mL^{st}&=&\mL^{st}_K+\mL^{st}_P \ , \nonumber \\
\mL^{st}_K&=&
\frac{T_{p-2}}{4\pi\alpha'\kappa^2}
\sqrt{\det \ba}
\tmD_t \tY^M
\tmD_t \tY_M
\nonumber \\
&+&\frac{\pi\alpha'T_{p-2}}{\kappa^2}
\sqrt{\det\ba}
\tmD_t \tA_\alpha
\bai^{\alpha\beta}_S\tmD_t \tA_\beta
\ , \nonumber \\
\mL_P&=&-\frac{\kappa^2T_{p-2}}{8}
\sqrt{\det\ba}
[2\pi\alpha'\tmJ^M\tmJ_M
+(2\pi\alpha')^{-1}\tmJ^i \ba^S_{ij}\tmJ^j]
\nonumber \\
\end{eqnarray}
where $\tY^M , M,N=1,\dots,10$ are
modes that parameterize transverse
positions of D(p-1)-brane, $\tA_\alpha$
are gauge fields living on its world
volume where the  world-volume of
D(p-1)-brane is parameterized with
coordinates $x^\alpha \ ,
\alpha,\beta=1,\dots,p-1$. Further,
$\tmD_t Y^M, \tmD_t \tA_\alpha, \tmJ_Y$
and $\tmJ_A$  are defined as
\begin{eqnarray}
\tmD_t \tY^M &=& \frac{1}{\tN}
\left(\partial_t \tY^M-\tN^\alpha \partial_\alpha
\tY^M\right) \ , \nonumber \\
\tmD_t A_\alpha
&=&\frac{1}{\tN}\left(
\tilde{E}_\alpha-\tN^\beta \tilde{F}_{\beta\alpha}
\right) \ , \nonumber \\
\tmJ_M(\bx)&=&
\frac{1}{\sqrt{\det\ba}}
\partial_\alpha[
\partial_\beta Y_M \bai^{\beta\alpha}_S
\sqrt{\det\ba}] \ ,
\nonumber \\
\tmJ^\alpha(\bx)&=&
\frac{2\pi\alpha'}{\sqrt{\det\ba}}
\partial_\beta[
\bai^{\beta\alpha}_A\sqrt{\det\ba}] \ ,
\nonumber \\
\end{eqnarray}
where
\begin{equation}
\ba_{\alpha\beta}=\partial_\alpha \tY^M\partial_\beta
\tY_M+2\pi\alpha'
\tilde{F}_{\alpha\beta} \ .
\end{equation}
Let us again consider the
action for non-relativistic
unstable Dp-brane (\ref{Ldi})
and analyze corresponding equation of
motion. Again, for simplicity we
consider the equation of motion for
the tachyon only and argue
the this equation of motion has
the solution
in the form of the
 tachyon kink \cite{Sen:2003tm}
\begin{eqnarray}\label{ans}
T&=&f(ax) \ , \quad  \frac{df}{du}\equiv
f'(u)>0 \ , \quad  \forall u \ , \quad  f(\pm \infty)=\pm \infty \ ,
\nonumber \\
Y^M&=&A_i=0 \ , \quad  N=\mathrm{const} \ ,
\quad N^i=0 \ , \nonumber \\
\end{eqnarray}
where $a$ is a parameter that should
be taken to infinity in the end.
As the first step we  determine
the equation of motion for $N$
\begin{eqnarray}\label{eqN}
& &\frac{\tau_{p-1}}{2\kappa^2}
V(T)\sqrt{\det \bA}\left[\frac{1}{2\pi\alpha'}
\mD_t Y^M\mD_t Y_M
+\mD_t T
\mD_t T+
2\pi\alpha'
\mD_t A_i\bAi^{ij}_S\mD_t A_j\right]+
\nonumber \\
&+&
\frac{\kappa^2\tau_{p-1}}{8}
V(T)\sqrt{\det\bA}
\left[2\pi\alpha'\mJ_M\mJ^M+
\mJ_T \mJ_T+(2\pi\alpha')^{-1}\mJ^i \bA^S_{ij}\mJ^j\right]=0
\nonumber \\
\end{eqnarray}
while the equation of motion for $N^i$
implies
\begin{eqnarray}\label{eqNi}
\frac{\tau_{p-1}}{2\kappa^2}
V(T)\sqrt{\det \bA}\left[\frac{1}{2\pi\alpha'}
\partial_i  Y^M\mD_t Y_M+
\partial_i T
\mD_t T+
 2\pi\alpha'  F_{ik}
\bAi^{kj}_S\mD_t A_j\right]
=0 \ .
 \nonumber \\
\end{eqnarray}
Now we see that
 (\ref{ans})
solves the equation (\ref{eqNi})
while for  $\mD_tT=0$
the equation (\ref{eqN}) reduces into
\begin{eqnarray}\label{eqNr}
\frac{\kappa^2\tau_{p-1}^2}{8}
V(T)\sqrt{\det\bA}
\mJ_T \mJ_T=0 \ .
\end{eqnarray}
We will argue below that the ansatz
(\ref{ans}) solves the equation (\ref{eqNr}) as
well. In fact, let us now
  analyze the equation of motion
for tachyon
\begin{eqnarray}
& &\frac{\tau_{p-1}}{2\kappa^2}
NV'\sqrt{\det\bA}
\left(\frac{1}{2\pi\alpha'}
\mD_t Y^M\mD_t Y_M+
\mD_t T\mD_t T+\mD_t A_i
\bAi^{ij}_S\mD_t A_j\right)-
\nonumber \\
&-&\frac{\pi\alpha'\tau_{p-1}}{\kappa^2}
\partial_i\left[NV\partial_j
T\bAi^{ji}_S
\sqrt{\det\bA}
\left(\frac{1}{2\pi\alpha'}
\mD_t Y^M\mD_t Y_M+
\mD_t T\mD_t T+\mD_t A_i
\bAi^{ij}_S\mD_t A_j\right)\right]
+\nonumber \\
&+&\frac{2\pi\alpha'\tau_{p-1}}
{\kappa^2}
\partial_k[NV(T)\sqrt{\det\bA}\mD_t A_i \bAi^{ik}_S
\partial_l T\bAi^{lj}\mD_tA_j]-
\nonumber \\
&-&\frac{\tau_{p-1}}
{\kappa^2}N\mD_t(
V(T)\sqrt{\det\bA}
\mD_t T)-\nonumber \\
&-&\frac{\kappa^2 N \tau_{p-1}}
{8}V'(T)\sqrt{\det\bA}
\left[\mJ_T\mJ_T+2\pi\alpha'
\mJ^M\mJ_M+\frac{1}{2\pi\alpha'}\mJ^i\bA_{ij}^S
\mJ^j\right]+\nonumber \\
&+&
\frac{\pi\alpha'\kappa^2
\tau_{p-1}}{4}
\partial_i\left[NV(T)
\partial_j T\bAi^{ji}
\sqrt{\det\bA}\left(\mJ_T \mJ_T+2\pi\alpha'
\mJ^M\mJ_M+\frac{1}{2\pi\alpha'}
\mJ^i\bA^S_{ij}\mJ^j\right)\right]-
\nonumber \\
&-&\frac{\kappa^2 N\tau_{p-1}^2}
{4}V(T)\sqrt{\det\bA}
(\frac{V''}{V}-\frac{V'^2}{V^2})\mJ_T-\nonumber \\
&-&\frac{\pi\alpha'\kappa^2 N\tau_{p-1}^2}
{2}V(T)\sqrt{\det\bA}\mJ_T
\frac{V'}{V^2\sqrt{\det\bA}}
\partial_i[\partial_j T
\bAi^{ji}\sqrt{\det\bA}]+
\nonumber \\
&+&\pi^2\alpha'^2\tau_{p-1}^2\kappa^2
\partial_i[N\mJ_T  \partial_j T
\bAi^{ji}\partial_k[V(T)\partial_l
T\bAi^{lk}\sqrt{\det\bA}]]-\nonumber \\
&-&\frac{\pi\alpha'\kappa^2 \tau_{p-1}^2}{2}
\partial_i[N\mJ_T]V'(T)
\partial_j T\bAi^{ji}
\sqrt{\det\bA}
+\frac{\pi\alpha'\kappa^2\tau_{p-1}^2}{2}
\partial_j\left[\partial_i[N \mJ_T]
V\bAi^{ji}\sqrt{\det\bA}\right]
-\nonumber \\
&-&\pi^2\alpha'^2\kappa^2\tau_{p-1}^2
\partial_k\left[\partial_i[N\mJ_T]V
 \partial_j T\bAi^{jk}
\partial_l T\bAi^{li}\sqrt{\det\bA}\right]+
\nonumber \\
&+&2\pi^2\alpha'^2\kappa^2\tau_{p-1}^2
\partial_k\left[
\partial_i[N\mJ_T] V
\partial_j T\bAi^{jk}\partial_lT \bAi^{li}
\sqrt{\det\bA}\right]+\dots=0 \ ,
\nonumber \\
\end{eqnarray}
where $"\dots"$ means terms that arise
from the variation of the expressions
$\mJ^M\mJ_M$ and $\mJ^i\bA_{ij}^S \mJ^j$.
We see that even in the flat background
the equation of motion for tachyon is
very complicated.

Note that for (\ref{ans})
the matrix $\bA$  and its determinant
are  equal to
\begin{equation}
\bA_{xx}=2\pi\alpha' a^2 f'^2 \ , \quad
\bA_{\alpha\beta}=\bA_{x\alpha}=0 \ , \quad
 \det{\bA}=\sqrt{2\pi\alpha'} af' \ .
\end{equation}
However the situation is much better
when we recognize that for (\ref{ans}) the
currents $\mJ$ vanish. In fact, $\mJ^M$ vanishes
since it contains derivative of $Y$ and
$\mJ^i$ vanishes since contains anti-symmetrization
of the matrix $\bA$ that for the ansatz
(\ref{ans}) is zero. Finally if we insert
the ansatz (\ref{ans}) into $\mJ_T$
we obtain
\begin{eqnarray}
\mJ_T=
\frac{V'}{V}
-\frac{2\pi\alpha'}{V \sqrt{2\pi\alpha'} af'}
\partial_x[\frac{V  af'\sqrt{2\pi \alpha' a^2 f'^2}}
{ 2\pi\alpha' a^2 f'^2}]=0 \ .
\nonumber \\
\end{eqnarray}
Further, as we argued above
$\mD_t T=0$ and hence   the
first fourth lines  vanish in the
equation of motion for tachyon. Then however using
the fact that $\mJ_T=0$ we find that the
variation of the potential term vanishes
as well and hence the ansatz (\ref{ans})
solves the tachyon equation of motion.

With analogy with familiar case of the
tachyon kink in superstring theory
\cite{Sen:2003tm} we interpret this
 solution as the stable non-relativistic
D(p-1)-brane.  To support further
this claim we should study the dynamics
of the fluctuations around this
tachyon kink exactly in the same
way as in \cite{Sen:2003tm}. Explicitly,
let us consider following ansatz for
fluctuations
\begin{eqnarray}\label{ansfluc}
T&=&T(a(x-t(x^\alpha))) \ , \quad Y^M(t,\bx)=\tY^M(t,\bx^\alpha)
\ , \nonumber \\
A_x&=&0 \ , \quad A_\alpha(t,\bx)=
\tA_\alpha(t,\bx^\alpha) \ ,
\nonumber \\
\end{eqnarray}
where $\bx^\alpha\equiv (x^1,\dots,x^{p-1})$. Then
we should insert (\ref{ansfluc}) to the
equations of motion that follow from the variation
of the action (\ref{Ldi}) and show that the equations
of motion are obeyed on condition that $\tY^M,\tA_\alpha$
obey the equations of motion that follow from the
variation of the action (\ref{Dp-1}) and hence their
dynamics really describes D(p-1)-brane. However
due to the complexity of the action (\ref{Ldi}) we
do not proceed in this way. Instead we show that
when we insert the ansatz (\ref{ansfluc}) into
(\ref{Ldi}) we reproduce the non-relativistic
D(p-1)-brane action.

The structure of the action (\ref{Ldi})
allows further important
simplification. By construction the
action (\ref{Ldi}) is invariant under
foliation-preserving diffeomorphism. On
the other hand we suggested the tachyon
fluctuation ansatz in the form
 $T=f(a(x-t(x^\alpha)))$. Then the invariance
of the action suggest to interpret $t$
as a parameter of diffeomorphism
transformation and hence it is
physically redundant and can be gauged
away.
 Then it is natural to
presume that the tachyon has the form
$T=f(a(x))$. However this result simplifies
the analysis considerably since then
 \begin{eqnarray}
 \bA_{\alpha\beta}&=& \ba_{\alpha\beta} \ , \quad
 \ba_{\alpha\beta}=\partial_\alpha \tY^M\partial_\beta
 \tY_M+2\pi\alpha'\tilde{F}_{\alpha\beta} \ , \nonumber \\
\bA_{x\beta}&=&\bA_{\alpha x}=0 \ , \quad
\bA_{xx}=a^2 f'^2  \ .
 \nonumber \\
 \end{eqnarray}
Inserting this result into the currents
$\mJ_T,\mJ_M$ and $\mJ^i$ we find
\begin{eqnarray}
\mJ_T=\frac{V'}{V}-\frac{(2\pi\alpha')^2}{V\sqrt{
\det \bA_{xx}\det \ba}}
\partial_x[\frac{V f \sqrt{
\det\ba}\sqrt{\bA_{xx}}}
{\bA_{xx}}]=0
\nonumber \\
\end{eqnarray}
and
\begin{eqnarray}
\mJ_M&=&\frac{1}{V(T)\sqrt{\bA_{xx}\det\ba}}
\partial_i[V(T)\partial_j
Y_M\bAi^{ji}_S\sqrt{\bA_{xx}\det\ba}]=
\nonumber \\
&=&\frac{1}{\sqrt{\det\ba}}
\partial_\alpha[\partial_\beta
\tY_M\bai^{\beta\alpha}_S\sqrt{\det\ba}]=\tmJ_M
\nonumber \\
\end{eqnarray}
and
\begin{eqnarray}
\mJ^i&=&-\frac{2\pi\alpha'}{V(T)\sqrt{\det\bA}}
\partial_i[V\bAi^{ij}\sqrt{\det\bA}]=
\nonumber \\
&=&-\frac{2\pi\alpha'}{\sqrt{\det\ba}}
\partial_\beta[\bai^{\beta\alpha}_A
\sqrt{\det\ba}]=\tmJ^i \ .
\nonumber \\
\end{eqnarray}
Using these results we easily obtain that
\begin{equation}
\mL_K+\mL_P=
\tau_{p-1}\sqrt{\bA_{xx}}V(f(ax))
(\tilde{\mL}_K+\tilde{\mL}_P) \ .
\end{equation}
Inserting this expression to the action
for an unstable  Dp-brane we obtain
\begin{eqnarray}
S=\int dt d^p\bx
N (\mL_K+\mL_P)&=&
\tau_{p-1}
\int d x af'(ax) V(f(ax))
\int dt d^{p-1}\bx N(\mL^{st}_K+
\mL^{st}_P)=
\nonumber \\
&=&T_{p-2}\int dt d^{p-1}\bx
N(\mL^{st}_K+\mL^{st}_P) \ ,
\nonumber \\
\end{eqnarray}
where by presumption
\begin{equation}
\tau_{p-1}\int dx af'(ax)V(af(x))=
\tau_{p-1}\int dm V(m)=T_{p-2}
\end{equation}
is the tension of D(p-1)-brane.
In other words we found that the action
for fluctuation around the tachyon
kink is the same as the action for
non-relativistic D(p-1)-brane.
\section{Conclusion}\label{seventh}
Let us outline results derived in this
paper. We found an action for stable
and unstable $p+1$ dimensional system
with anisotropic scaling that has the
property that the potential is
proportional to the variation of stable
or unstable Euclidean Dp-brane action.
We extended the symmetries of these
$p+1$ dimensional theories in order to
be invariant under folliation
preserving diffeomorphism and under
space-time dependent gauge
transformations. Then we developed the
Hamiltonian formalism for these
theories and we also studied the
tachyon kink  on the world-volume of
$p+1$ dimensional unstable theory. We
argued that the tachyon kink
corresponds to the stable $p$
dimensional non-relativistic theory.

Even if it is an open question how
these theories can be embedded into
superstring theories we mean that they
provide an interesting new class of
non-relativistic theories that can be
studied further. \vskip .2in \noindent
{\bf Acknowledgements:} This work was
 supported by the Czech
Ministry of Education under Contract
No. MSM 0021622409. I would like also
thank to Max Planck Institute at Golm
for its kind hospitality during my work on
this project.

\newpage


\begin{thebibliography}{20}








\bibitem{Horava:2009if}
  P.~Horava,
\emph{``Spectral Dimension
of Spacetime in Quantum Gravity at a Lifshitz Point,''}
  arXiv:0902.3657 [hep-th].

\bibitem{Horava:2009uw}
  P.~Horava,
\emph{``Quantum Gravity at a Lifshitz Point,''}
  arXiv:0901.3775 [hep-th].

\bibitem{Horava:2008ih}
  P.~Horava,
\emph{``Membranes at Quantum Criticality,''}
  arXiv:0812.4287 [hep-th].

\bibitem{Horava:2008jf}
  P.~Horava,
\emph{``Quantum Criticality
 and Yang-Mills Gauge Theory,''}
  arXiv:0811.2217 [hep-th].












\bibitem{Myung:2009va}
  Y.~S.~Myung,
\emph{``Thermodynamics of
black holes in the deformed Ho\v{r}ava-Lifshitz
gravity,''}
  arXiv:0905.0957 [hep-th].

\bibitem{Ghodsi:2009rv}
  A.~Ghodsi,
\emph{``Toroidal solutions in Horava Gravity,''}
  arXiv:0905.0836 [hep-th].

\bibitem{Cai:2009qs}
  R.~G.~Cai, L.~M.~Cao and N.~Ohta,
\emph{``Thermodynamics of
Black Holes in Horava-Lifshitz Gravity,''}
  arXiv:0905.0751 [hep-th].

\bibitem{Rama:2009px}
  S.~K.~Rama,
\emph{``Anisotropic Cosmology
 and (Super)Stiff Matter in Ho\v{r}ava's Gravity
Theory,''}
  arXiv:0905.0700 [hep-th].

\bibitem{Kehagias:2009is}
  A.~Kehagias and K.~Sfetsos,
\emph{``The black hole and
 FRW geometries of non-relativistic gravity,''}
  arXiv:0905.0477 [hep-th].

\bibitem{Nishioka:2009iq}
  T.~Nishioka,
\emph{``Horava-Lifshitz Holography,''}
  arXiv:0905.0473 [hep-th].










\bibitem{Gao:2009er}
  C.~Gao,
\emph{``Modified gravity in
 Arnowitt-Deser-Misner formalism,''}
  arXiv:0905.0310 [astro-ph.CO].

\bibitem{Orlando:2009en}
  D.~Orlando and S.~Reffert,
\emph{``On the Renormalizability
of Horava-Lifshitz-type Gravities,''}
  arXiv:0905.0301 [hep-th].

\bibitem{Cai:2009dx}
  R.~G.~Cai, B.~Hu and H.~B.~Zhang,
\emph{``Dynamical Scalar
 Degree of Freedom in Horava-Lifshitz Gravity,''}
  arXiv:0905.0255 [hep-th].

\bibitem{Myung:2009dc}
  Y.~S.~Myung and Y.~W.~Kim,
\emph{``Thermodynamics of
Ho\v{r}ava-Lifshitz black holes,''}
  arXiv:0905.0179 [hep-th].







\bibitem{Mukohyama:2009zs}
  S.~Mukohyama, K.~Nakayama, F.~Takahashi and S.~Yokoyama,
\emph{``Phenomenological Aspects of
Horava-Lifshitz Cosmology,''}
  arXiv:0905.0055 [hep-th].

\bibitem{Chen:2009ka}
  B.~Chen and Q.~G.~Huang,
\emph{``Field Theory at a Lifshitz
Point,''}
  arXiv:0904.4565 [hep-th].


\bibitem{Sotiriou:2009gy}
  T.~Sotiriou, M.~Visser and S.~Weinfurtner,
\emph{``Phenomenologically viable
Lorentz-violating quantum gravity,''}
  arXiv:0904.4464 [hep-th].


\bibitem{Colgain:2009fe}
  E.~O.~Colgain and H.~Yavartanoo,
\emph{``Dyonic solution of
Horava-Lifshitz Gravity,''}
  arXiv:0904.4357 [hep-th].


\bibitem{Gao:2009bx}
  X.~Gao,
\emph{``Cosmological Perturbations and
Non-Gaussianities in
Ho\v{r}ava-Lifshitz
  Gravity,''}
  arXiv:0904.4187 [hep-th].





\bibitem{Piao:2009ax}
  Y.~S.~Piao,
 \emph{"Primordial
 Perturbation in Horava-Lifshitz
 Cosmology,''}
  arXiv:0904.4117 [hep-th].

\bibitem{Volovik:2009av}
  G.~E.~Volovik,
\emph{``z=3 Lifshitz-Horava model and
Fermi-point scenario of emergent
gravity,''}
  arXiv:0904.4113 [gr-qc].


\bibitem{Cai:2009ar}
  R.~G.~Cai, Y.~Liu and Y.~W.~Sun,
\emph{``On the z=4 Horava-Lifshitz
Gravity,''}
  arXiv:0904.4104 [hep-th].






\bibitem{Cai:2009pe}
  R.~G.~Cai, L.~M.~Cao and N.~Ohta,
\emph{``Topological Black Holes in Horava-Lifshitz Gravity,''}
  arXiv:0904.3670 [hep-th].


\bibitem{Nastase:2009nk}
  H.~Nastase,
\emph{``On IR solutions in Horava gravity theories,''}
  arXiv:0904.3604 [hep-th].

\bibitem{Izawa:2009ne}
  K.~I.~Izawa,
\emph{``A Note on Quasi-Riemannian Gravity with Higher Derivatives,''}
  arXiv:0904.3593 [hep-th].


\bibitem{Brandenberger:2009yt}
  R.~Brandenberger,
\emph{``Matter Bounce in Horava-Lifshitz Cosmology,''}
  arXiv:0904.2835 [hep-th].

\bibitem{Mukohyama:2009gg}
  S.~Mukohyama,
\emph{``Scale-invariant cosmological perturbations
from Horava-Lifshitz gravity
  without inflation,''}
  arXiv:0904.2190 [hep-th].

\bibitem{Lu:2009em}
  H.~Lu, J.~Mei and C.~N.~Pope,
\emph{``Solutions to Horava Gravity,''}
  arXiv:0904.1595 [hep-th].





\bibitem{Kiritsis:2009sh}
  E.~Kiritsis and G.~Kofinas,
\emph{``Horava-Lifshitz Cosmology,''}
  arXiv:0904.1334 [hep-th].

\bibitem{Calcagni:2009ar}
  G.~Calcagni,
\emph{``Cosmology of the Lifshitz universe,''}
  arXiv:0904.0829 [hep-th].

\bibitem{Takahashi:2009wc}
  T.~Takahashi and J.~Soda,
\emph{``Chiral Primordial
Gravitational Waves from a Lifshitz Point,''}
  arXiv:0904.0554 [hep-th].

\bibitem{Jenkins:2009un}
  A.~Jenkins,
\emph{``Constraints on emergent gravity,''}
  arXiv:0904.0453 [gr-qc].



\bibitem{Visser:2009fg}
  M.~Visser,
\emph{``Lorentz symmetry
 breaking as a quantum field theory regulator,''}
  arXiv:0902.0590 [hep-th].






\bibitem{Kluson:2009sm}
  J.~Kluson,
\emph{``Branes at Quantum Criticality,''}
  arXiv:0904.1343 [hep-th].








\bibitem{Ardonne:2003wa}
  E.~Ardonne, P.~Fendley and E.~Fradkin,
\emph{``Topological order and conformal
quantum critical points,''}
  Annals Phys.\  {\bf 310} (2004) 493
  [arXiv:cond-mat/0311466].


\bibitem{Blau:2009gd}
  M.~Blau, J.~Hartong and B.~Rollier,
\emph{``Geometry of Schroedinger
 Space-Times, Global Coordinates, and Harmonic
Trapping,''}
  arXiv:0904.3304 [hep-th].



\bibitem{Volovich:2009yh}
  A.~Volovich and C.~Wen,
\emph{``Correlation Functions
 in Non-Relativistic Holography,''}
  arXiv:0903.2455 [hep-th].



\bibitem{Pal:2009np}
  S.~S.~Pal,
\emph{``Non-relativistic supersymmetric Dp branes,''}
  arXiv:0904.3620 [hep-th].



\bibitem{Son:2008ye}
  D.~T.~Son,
\emph{``Toward an AdS/cold atoms
correspondence: a geometric realization
of the Schroedinger symmetry,''}
  Phys.\ Rev.\  D {\bf 78}, 046003 (2008)
  [arXiv:0804.3972 [hep-th]].

\bibitem{Balasubramanian:2008dm}
  K.~Balasubramanian and J.~McGreevy,
\emph{``Gravity duals for
non-relativistic CFTs,''}
  Phys.\ Rev.\ Lett.\  {\bf 101}, 061601 (2008)
  [arXiv:0804.4053 [hep-th]].

\bibitem{Goldberger:2008vg}
  W.~D.~Goldberger,
\emph{``AdS/CFT duality for
non-relativistic field theory,''}
  arXiv:0806.2867 [hep-th].

\bibitem{Barbon:2008bg}
  J.~L.~B.~Barbon and C.~A.~Fuertes,
\emph{``On the spectrum of
nonrelativistic AdS/CFT,''}
  JHEP {\bf 0809}, 030 (2008)
  [arXiv:0806.3244 [hep-th]].

\bibitem{Wen:2008hi}
  W.~Y.~Wen,
\emph{``AdS/NRCFT for the (super)
Calogero model,''}
  arXiv:0807.0633 [hep-th].

\bibitem{Herzog:2008wg}
  C.~P.~Herzog, M.~Rangamani and S.~F.~Ross,
\emph{``Heating up Galilean
holography,''}
  JHEP {\bf 0811}, 080 (2008)
  [arXiv:0807.1099 [hep-th]].

\bibitem{Maldacena:2008wh}
  J.~Maldacena, D.~Martelli and Y.~Tachikawa,
\emph{``Comments on string theory
backgrounds with non-relativistic
conformal symmetry,''}
  JHEP {\bf 0810}, 072 (2008)
  [arXiv:0807.1100 [hep-th]].

\bibitem{Adams:2008wt}
  A.~Adams, K.~Balasubramanian and J.~McGreevy,
\emph{``Hot Spacetimes for Cold
Atoms,''}
  JHEP {\bf 0811}, 059 (2008)
  [arXiv:0807.1111 [hep-th]].



\bibitem{Minic:2008xa}
  D.~Minic and M.~Pleimling,
\emph{``Non-relativistic AdS/CFT and
Aging/Gravity Duality,''}
  arXiv:0807.3665 [cond-mat.stat-mech].

\bibitem{Chen:2008ad}
  J.~W.~Chen and W.~Y.~Wen,
\emph{``Shear Viscosity of a
Non-Relativistic Conformal Gas in Two
Dimensions,''}
  arXiv:0808.0399 [hep-th].

\bibitem{Colgain:2009wm}
  E.~O.~Colgain and H.~Yavartanoo,
\emph{``NR $CFT_3$ duals in M-theory,''}
  arXiv:0904.0588 [hep-th].




\bibitem{Bagchi:2009my}
  A.~Bagchi and R.~Gopakumar,
\emph{``Galilean Conformal Algebras and
AdS/CFT,''}
  arXiv:0902.1385 [hep-th].

\bibitem{Alishahiha:2009hg}
  M.~Alishahiha and A.~Ghodsi,
\emph{``Non-relativistic D3-brane in
the presence of higher derivative
  corrections,''}
  arXiv:0901.3431 [hep-th].


\bibitem{Donos:2009en}
  A.~Donos and J.~P.~Gauntlett,
\emph{``Supersymmetric solutions for
non-relativistic holography,''}
  arXiv:0901.0818 [hep-th].


\bibitem{Pal:2009yp}
  S.~S.~Pal,
\emph{``Anisotropic gravity solutions
in AdS/CMT,''}
  arXiv:0901.0599 [hep-th].

\bibitem{Danielsson:2008gi}
  U.~H.~Danielsson and L.~Thorlacius,
\emph{``Black holes in asymptotically
Lifshitz spacetime,''}
  arXiv:0812.5088 [hep-th].

\bibitem{Taylor:2008tg}
  M.~Taylor,
\emph{``Non-relativistic holography,''}
  arXiv:0812.0530 [hep-th].

\bibitem{Adams:2008zk}
  A.~Adams, A.~Maloney, A.~Sinha and S.~E.~Vazquez,
\emph{``1/N Effects in Non-Relativistic
Gauge-Gravity Duality,''}
  arXiv:0812.0166 [hep-th].

\bibitem{Akhavan:2008ep}
  A.~Akhavan, M.~Alishahiha, A.~Davody and A.~Vahedi,
\emph{``Non-relativistic CFT and
Semi-classical Strings,''}
  arXiv:0811.3067 [hep-th].

\bibitem{Rangamani:2008gi}
  M.~Rangamani, S.~F.~Ross, D.~T.~Son and E.~G.~Thompson,
 \emph{"Conformal non-relativistic
 hydrodynamics from gravity,''}
  JHEP {\bf 0901} (2009) 075
  [arXiv:0811.2049 [hep-th]].

\bibitem{Mazzucato:2008tr}
  L.~Mazzucato, Y.~Oz and S.~Theisen,
\emph{``Non-relativistic Branes,''}
  arXiv:0810.3673 [hep-th].

\bibitem{Schvellinger:2008bf}
  M.~Schvellinger,
\emph{``Kerr-AdS black holes and
non-relativistic conformal QM theories
in diverse dimensions,''}
  JHEP {\bf 0812} (2008) 004
  [arXiv:0810.3011 [hep-th]].


\bibitem{Sachdev:2008ba}
  S.~Sachdev and M.~Mueller,
\emph{``Quantum criticality and black
holes,''}
  arXiv:0810.3005 [cond-mat.str-el].


\bibitem{Hartnoll:2008rs}
  S.~A.~Hartnoll and K.~Yoshida,
\emph{``Families of IIB duals for
nonrelativistic CFTs,''}
  JHEP {\bf 0812} (2008) 071
  [arXiv:0810.0298 [hep-th]].

\bibitem{Lin:2008pi}
  F.~L.~Lin and S.~Y.~Wu,
\emph{``Non-relativistic Holography and
Singular Black Hole,''}
  arXiv:0810.0227 [hep-th].

\bibitem{Yamada:2008if}
  D.~Yamada,
\emph{``Thermodynamics of Black Holes
in Schroedinger Space,''}
  arXiv:0809.4928 [hep-th].

\bibitem{Duval:2008jg}
  C.~Duval, M.~Hassaine and P.~A.~Horvathy,
\emph{``The geometry of Schr\'odinger
symmetry in gravity
  background/non-relativistic CFT,''}
  arXiv:0809.3128 [hep-th].

\bibitem{Kovtun:2008qy}
  P.~Kovtun and D.~Nickel,
\emph{``Black holes and
non-relativistic quantum systems,''}
  Phys.\ Rev.\ Lett.\  {\bf 102} (2009) 011602
  [arXiv:0809.2020 [hep-th]].


\bibitem{Kachru:2008yh}
  S.~Kachru, X.~Liu and M.~Mulligan,
\emph{``Gravity Duals of Lifshitz-like
Fixed Points,''}
  Phys.\ Rev.\  D {\bf 78} (2008) 106005
  [arXiv:0808.1725 [hep-th]].









\bibitem{Lee:2009mm}
  K.~M.~Lee, S.~Lee and S.~Lee,
\emph{``Nonrelativistic Superconformal
M2-Brane Theory,''}
  arXiv:0902.3857 [hep-th].

\bibitem{Galajinsky:2009dw}
  A.~Galajinsky and I.~Masterov,
\emph{``Remark on quantum mechanics
with N=2 superconformal Galilean
symmetry,''}
  arXiv:0902.2910 [hep-th].

\bibitem{Nakayama:2009cz}
  Y.~Nakayama, M.~Sakaguchi and K.~Yoshida,
\emph{``Non-Relativistic M2-brane Gauge
Theory and New Superconformal
Algebra,''}
  arXiv:0902.2204 [hep-th].


\bibitem{Nakayama:2008qm}
  Y.~Nakayama,
\emph{``Index for Non-relativistic
Superconformal Field Theories,''}
  JHEP {\bf 0810} (2008) 083
  [arXiv:0807.3344 [hep-th]].



\bibitem{Gomis:2004pw}
  J.~Gomis, K.~Kamimura and P.~K.~Townsend,
\emph{``Non-relativistic
superbranes,''}
  JHEP {\bf 0411} (2004) 051
  [arXiv:hep-th/0409219].














\bibitem{Brugues:2004an}
  J.~Brugues, T.~Curtright, J.~Gomis and L.~Mezincescu,
\emph{``Non-relativistic strings and
branes as non-linear realizations of
Galilei groups,''}
  Phys.\ Lett.\  B {\bf 594} (2004) 227
  [arXiv:hep-th/0404175].

\bibitem{Kluson:2006xi}
  J.~Kluson,
\emph{``Non-relativistic non-BPS
Dp-brane,''}
  Nucl.\ Phys.\  B {\bf 765} (2007) 185
  [arXiv:hep-th/0610073].

\bibitem{Gomis:2006wu}
  J.~Gomis, K.~Kamimura and P.~C.~West,
\emph{``Diffeomorphism, kappa
transformations and the theory of
non-linear realisations,''}
  JHEP {\bf 0610} (2006) 015
  [arXiv:hep-th/0607104].

\bibitem{Gomis:2006xw}
  J.~Gomis, K.~Kamimura and P.~C.~West,
 \emph{"The
 construction of brane and
  superbrane actions using non-linear
  realisations,''}
  Class.\ Quant.\ Grav.\  {\bf 23} (2006) 7369
  [arXiv:hep-th/0607057].

\bibitem{Sakaguchi:2006pg}
  M.~Sakaguchi and K.~Yoshida,
\emph{``Non-relativistic AdS branes and
Newton-Hooke superalgebra,''}
  JHEP {\bf 0610} (2006) 078
  [arXiv:hep-th/0605124].

\bibitem{Gomis:2005bj}
  J.~Gomis, F.~Passerini, T.~Ramirez and A.~Van Proeyen,
\emph{``Non relativistic Dp branes,''}
  JHEP {\bf 0510} (2005) 007
  [arXiv:hep-th/0507135].

\bibitem{Gomis:2005pg}
  J.~Gomis, J.~Gomis and K.~Kamimura,
\emph{``Non-relativistic superstrings:
A new soluble sector of AdS(5) x
S**5,''}
  JHEP {\bf 0512} (2005) 024
  [arXiv:hep-th/0507036].

\bibitem{Seiberg:1999vs}
  N.~Seiberg and E.~Witten,
\emph{``String theory and noncommutative geometry,''}
  JHEP {\bf 9909} (1999) 032
  [arXiv:hep-th/9908142].

\bibitem{Kutasov:2003er}
  D.~Kutasov and V.~Niarchos,
\emph{``Tachyon effective
 actions in open string theory,''}
  Nucl.\ Phys.\  B {\bf 666} (2003) 56
  [arXiv:hep-th/0304045].

\bibitem{Sen:1999md}
  A.~Sen,
\emph{``Supersymmetric
 world-volume action for non-BPS D-branes,''}
  JHEP {\bf 9910}, 008 (1999)
  [arXiv:hep-th/9909062].

\bibitem{Garousi:2000tr}
  M.~R.~Garousi,
\emph{``Tachyon couplings
 on non-BPS D-branes and Dirac-Born-Infeld action,''}
  Nucl.\ Phys.\  B {\bf 584}, 284 (2000)
  [arXiv:hep-th/0003122].

\bibitem{Bergshoeff:2000dq}
  E.~A.~Bergshoeff, M.~de Roo, T.~C.~de Wit, E.~Eyras and S.~Panda,
\emph{``T-duality and
actions for non-BPS D-branes,''}
  JHEP {\bf 0005}, 009 (2000)
  [arXiv:hep-th/0003221].

\bibitem{Kluson:2000iy}
  J.~Kluson,
\emph{``Proposal for non-BPS D-brane action,''}
  Phys.\ Rev.\  D {\bf 62}, 126003 (2000)
  [arXiv:hep-th/0004106].

\bibitem{Sen:2003tm}
  A.~Sen,
\emph{``Dirac-Born-Infeld
action on the tachyon kink and vortex,''}
  Phys.\ Rev.\  D {\bf 68}, 066008 (2003)
  [arXiv:hep-th/0303057].


\bibitem{Sen:2004nf}
  A.~Sen,
\emph{``Tachyon dynamics in open string theory,''}
  Int.\ J.\ Mod.\ Phys.\  A {\bf 20} (2005) 5513
  [arXiv:hep-th/0410103].



\bibitem{Kluson:2005fj}
  J.~Kluson,
\emph{``Tachyon kink on
non-BPS Dp-brane in the general background,''}
  JHEP {\bf 0510} (2005) 076
  [arXiv:hep-th/0508239].







\end{thebibliography}
\end{document}